\documentclass[referee,sn-nature,pdflatex,sn-mathphys]{sn-jnl}

\usepackage{graphicx}%
\usepackage{multirow}%
\usepackage{amsmath,amssymb,amsfonts}%
\usepackage{amsthm}%
\usepackage{mathrsfs}%
\usepackage[title]{appendix}%
\usepackage{xcolor}%
\usepackage{textcomp}%
\usepackage{manyfoot}%
\usepackage{booktabs}%
\usepackage{algorithm}%
\usepackage{algorithmicx}%
\usepackage{algpseudocode}%
\usepackage{listings}%
\usepackage{siunitx}

\usepackage{geometry}
\usepackage{nicefrac}

\jyear{2024}%

\theoremstyle{thmstyleone}%
\theoremstyle{thmstyletwo}%

\theoremstyle{thmstylethree}%

\begin{document}

\title[Marchiori et al.]{Imaging magnetic spiral phases, skyrmion clusters, and skyrmion displacements at the surface of bulk Cu$_2$OSeO$_3$}


\author[1]{\fnm{E.} \sur{Marchiori}}

\author[1]{\fnm{G.} \sur{Romagnoli}}

\author[1,2]{\fnm{L.} \sur{Schneider}}

\author[1]{\fnm{B.} \sur{Gross}}

\author[3,4]{\fnm{P.} \sur{Sahafi}}

\author[3,4]{\fnm{A.} \sur{Jordan}}

\author[3,4]{\fnm{R.} \sur{Budakian}}

\author[5,6]{\fnm{P. R.} \sur{Baral}}

\author[6]{\fnm{A.} \sur{Magrez}}

\author[5]{\fnm{J. S.} \sur{White}}

\author*[1,2]{\fnm{M.} \sur{Poggio}}\email{martino.poggio@unibas.ch}

\affil[1]{\orgdiv{Department of Physics}, \orgname{University of Basel}, \orgaddress{\country{Switzerland}}}

\affil[2]{\orgdiv{Swiss Nanoscience Institute}, \orgname{University of Basel}, \orgaddress{\country{Switzerland}}}

\affil[3]{\orgdiv{Department of Physics and Astronomy}, \orgname{University of Waterloo}, \orgaddress{\country{Canada}}}

\affil[4]{\orgdiv{Institute for Quantum Computing}, \orgname{University of Waterloo}, \orgaddress{\country{Canada}}}

\affil[5]{\orgdiv{Laboratory for Neutron Scattering and Imaging}, \orgname{Paul Scherrer Institute}, \orgaddress{\country{Switzerland}}}

\affil[6]{\orgdiv{Institute of Physics}, \orgname{\'Ecole Polytechnique F\'ed\'erale de Lausanne}, \orgaddress{\country{Switzerland}}}



\abstract{Surfaces -- by breaking bulk symmetries, introducing roughness, or hosting defects -- can significantly influence magnetic order in magnetic materials. Determining their effect on the complex nanometer-scale phases present in certain non-centrosymmetric magnets is an outstanding problem requiring high-resolution magnetic microscopy. Here, we use scanning SQUID-on-tip microscopy to image the surface of bulk Cu$_2$OSeO$_3$ at low temperature and in a magnetic field applied along $\left\langle100\right\rangle$. Real-space maps measured as a function of applied field reveal the microscopic structure of the magnetic phases and their transitions. In low applied field, we observe a magnetic texture consistent with an in-plane stripe phase, pointing to the existence of a distinct surface state. In the low-temperature skyrmion phase, the surface is populated by clusters of disordered skyrmions, which emerge from rupturing domains of the tilted spiral phase. Furthermore, we displace individual skyrmions from their pinning sites by applying an electric potential to the scanning probe, thereby demonstrating local skyrmion control at the surface of a magnetoelectric insulator.}




\maketitle

\section{Introduction}\label{intro}

Surfaces have a profound impact on a number of physical phenomena by breaking the uniformity of the bulk and introducing the complexity associated with boundaries, asymmetry, and imperfection. Because they form a material’s interface with its environment, they often determine the nature of this interaction. In magnetic materials, the magnetic order at the surface -- especially in devices used for high-density information storage or processing -- determines performance and functionality. This notion also applies to magnetic skyrmions~\cite{bogdanov_physical_2020}. Despite being found in only a few materials meeting special symmetry requirements~\cite{tokura_magnetic_2021}, over the last 15 years, these discrete nanometer-scale magnetization configurations have captivated the attention of researchers, because of their non-trivial topology, nanometer-scale size, and the possibility of electrically manipulating them. All of these properties make skyrmions promising for information storage and processing applications~\cite{fert_magnetic_2017,song_skyrmion-based_2020}. As a result, researchers have extensively studied the magnetic behavior of skyrmion-hosting materials both in bulk and low-dimensional forms via a variety of techniques. Although it is known that the surface of such materials can exhibit magnetic order not present in the bulk~\cite{rybakov_new_2015,rybakov_new_2016,zhang_reciprocal_2018,zheng_experimental_2018,zhang_direct_2018,zhang_robust_2020,burn_periodically_2021,turnbull_x-ray_2022,xie_observation_2023,jin_evolution_2023}, studies of how magnetic phases change upon encountering the surface remain scarce. Likewise, microscopy of the surface of bulk materials, which can yield otherwise inaccessible information on underlying microscopic configurations and domain structure, is in short supply.

Among skyrmion-hosting materials, the insulating cubic helimagnet Cu$_2$OSeO$_3$ is of particular interest because of its magnetoelectric coupling, which allows the control of skyrmions via electric fields. Unlike control via electric current, which dissipates energy through Joule heating, control via electric fields can be carried out with negligible losses, which -- in principle -- allows for low-power device design. A number of experiments have exploited the coupling of an external electric field with the emergent electric polarization in multiferroic materials to tune the relative energies of competing magnetic configurations. In this way, for example, electric field has been used to rotate~\cite{white_electric-field-induced_2014} or even create~\cite{okamura_transition_2016,okamura_directional_2017,kruchkov_direct_2018,white_electric-field-driven_2018,huang_situ_2018,wilson_measuring_2019,huang_electric_2022} the skyrmion lattice phase in Cu$_2$OSeO$_3$. Although Mochizuki and Watanabe proposed a scheme for locally creating individual skyrmions in thin multiferroic films using a charged tip~\cite{mochizuki_writing_2015}, such control has only been demonstrated in epitaxial ferromagnetic films via scanning tunneling microscopy~\cite{romming_writing_2013,hsu_electric-field-driven_2017}. 

In addition to magnetic skyrmions, Cu$_2$OSeO$_3$ also hosts other periodically modulated magnetic phases. Like in other non-centrosymmetric cubic crystals, such as MnSi, Mn$_{1-x}$Co$_x$Si, MnGe, and FeGe, the lack of inversion symmetry induces an antisymmetric exchange interaction known as the Dzyaloshinskii-Moriya interaction (DMI), which is proportional to the spin-orbit coupling. This interaction and a common hierarchy of magnetic energy scales in these materials result in a similar set of modulated magnetic phases, including helical (H) and skyrmion lattice phases, and the corresponding transitions between them. Cu$_2$OSeO$_3$, however, deviates from this universal behavior, exhibiting a tilted spiral (TS) phase and a low-temperature skyrmion (LTS) phase, arising from the competition between anisotropic magnetic interactions. These states appear only under magnetic fields applied along $\left\langle100\right\rangle$, highlighting the role of cubic magnetocrystalline anisotropy and anisotropic exchange interaction for their stabilization~\cite{crisanti_tilted_2023,baral_direct_2023}. Until now, observations of these phases have been based on small angle neutron scattering (SANS), as well as measurements of magnetization and susceptibility~\cite{chacon_observation_2018,qian_new_2018,halder_thermodynamic_2018,bannenberg_multiple_2019}. Although these techniques provide a wealth of information on the magnetic phases in the bulk, they do not yield their real-space configuration or their properties near the surface. In particular, the nucleation and structure of the LTS phase remain open questions~\cite{leonov_topological_2022}. Real-space imaging at low-temperature with field applied along $\langle100\rangle$ has only been carried out via Lorentz transmission electron microscopy (LTEM) on a thin lamella of Cu$_2$OSeO$_3$, whose reduced dimensionality and milling by focused ion beam alter intrinsic properties. The surface of bulk Cu$_2$OSeO$_3$ has been imaged via magnetic force microscopy (MFM) in the high-temperature skyrmion phase~\cite{milde_heuristic_2016} and at low temperatures for applied fields close to the $\left\langle110\right\rangle$ axis, where the TS and LTS phases are not present~\cite{milde_field-induced_2020}.


\begin{figure}[t]%
\centering
\includegraphics[width=0.8\textwidth]{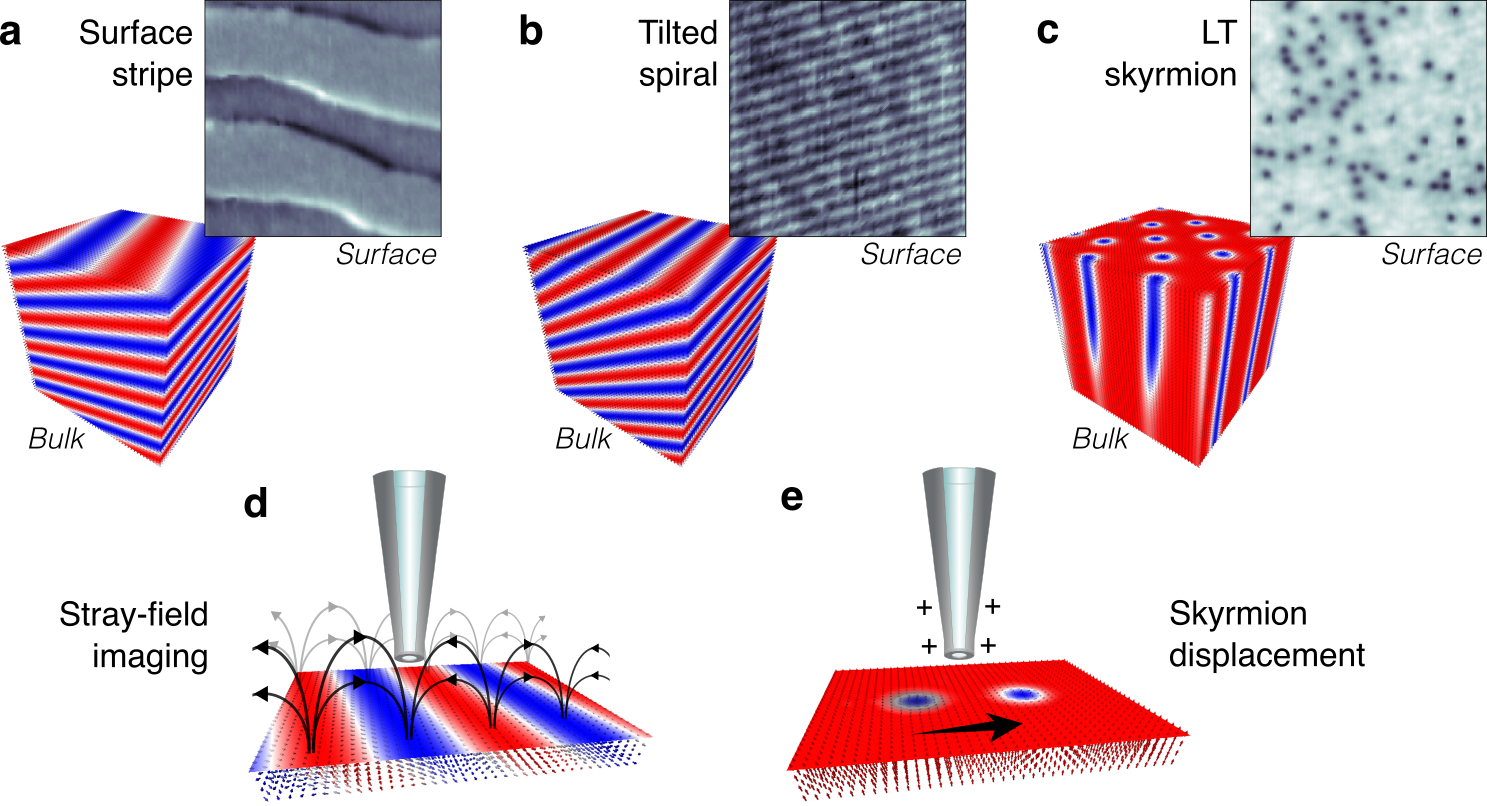}
\caption{Illustrative summary. \textbf{a},~Representation of the bulk magnetization configuration of a helical or conical phase slightly misaligned with the surface and the corresponding measured stray-field of surface stripes. \textbf{b},~Representation of the bulk magnetization and measured stray-field image of a tilted spiral phase. \textbf{c},~Representation of the bulk magnetization and measured stray-field of a low-temperature skyrmion phase. \textbf{d},Illustration of stray-field imaging with a SQUID-on-tip. \textbf{e},~Illustration of skyrmion displacement by a charged SQUID-on-tip.}\label{fig0_summary}
\end{figure}

Here, as shown schematically in Fig.~\ref{fig0_summary}, we image the stray magnetic field at the surface of bulk Cu$_2$OSeO$_3$ at low temperature and with a magnetic field applied along $\left\langle100\right\rangle$.  We use nanometer-scale scanning superconducting quantum interference device (SQUID) microscopy (SSM) to map the real-space configuration of the TS and LTS phases. Images taken as a function of applied magnetic field capture how the transitions between these states unfold, including the rupture of TS domains into clusters of skyrmions and the coexistence of field-polarized (FP), TS, and LTS domains.  In low magnetic fields, high-resolution MFM reveals the transition from the conical (C) phase to the multi-domain H phase. In this regime, we also observe patterns consistent with in-plane magnetic stripes, which -- because of their absence in bulk measurements -- may constitute a distinct surface state. Finally, images of the LTS phase reveal clusters of disordered skyrmions dominated by random pinning. By applying a localized electric field between the SQUID-on-tip probe and the sample, we displace skyrmions along the surface from one pinning site to another. Understanding this mechanism and the role of pinning or quenched disorder at the surface of skyrmion hosting materials is crucial for any applications relying on their manipulation~\cite{reichhardt_statics_2022}.

\section{Results}
\subsection{Low-temperature magnetic phases}\label{sec1}

We map the stray magnetic field just above the $\left(001\right)$ surface of a Cu$_2$OSeO$_3$ crystal at 5~K (see see \ref{AppA1}). Specifically, we measure the component of the field perpendicular to the surface, $B_z$. By oscillating the probe along the $y$-axis, we also measure $B_z^{ac} \propto d B_z / d y$, which offers a higher spatial resolution and magnetic sensitivity (see \ref{AppA2}). Maps of both $B_z(x,y)$ and $B_z^{ac}(x,y)$ measured at this location show the progression of magnetic states present at the surface as a function of magnetic field $H$ applied along $\left[001\right]$ and are representative of images taken at various positions on the surface of this, as well as of a second crystal.

\begin{figure}[t]%
\centering
\includegraphics[width=1.0\textwidth]{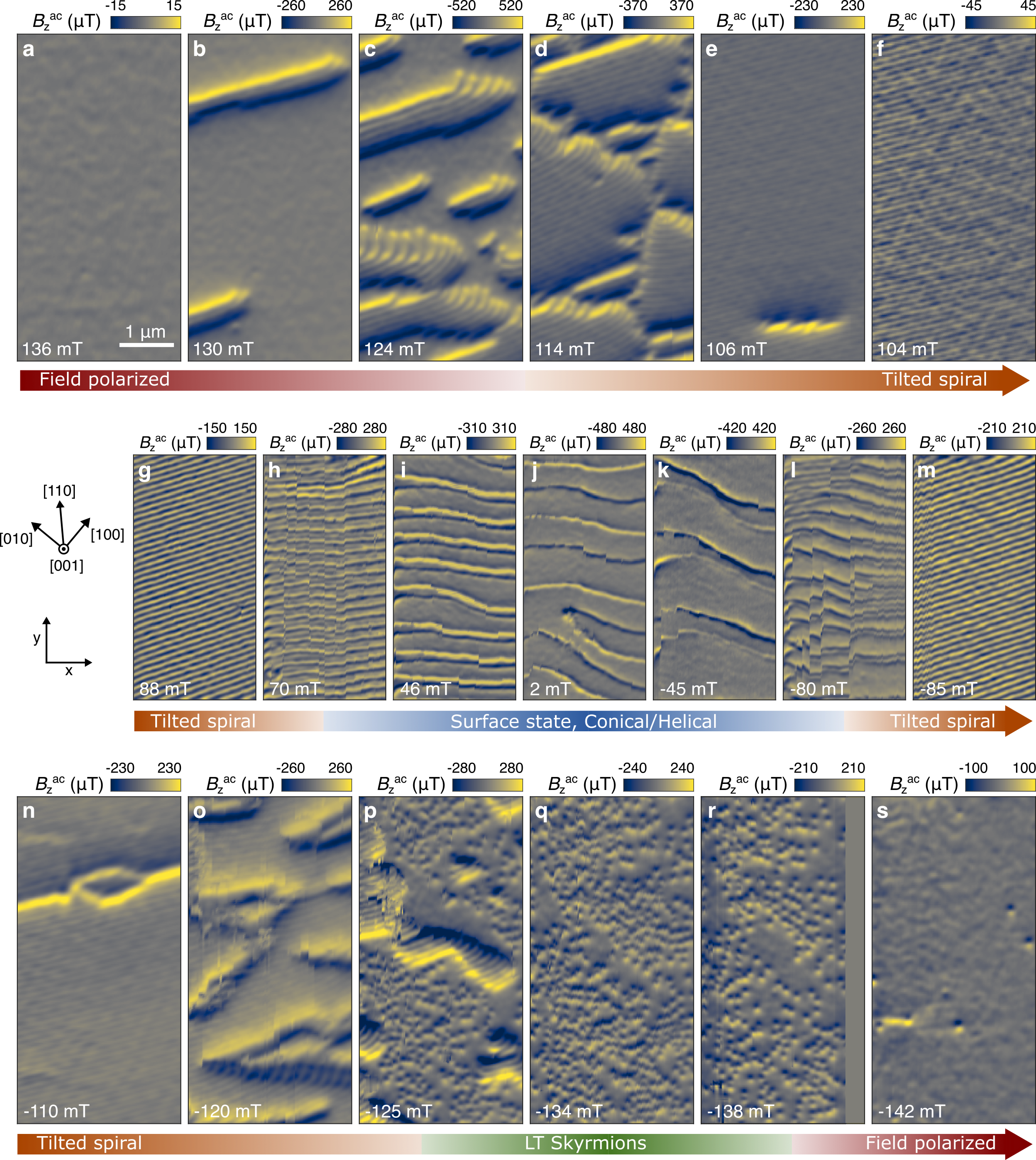}
\caption{SSM of magnetic phases at the surface of Cu$_2$OSeO$_3$. \textbf{a}-\textbf{s}~Images of $B_z^{ac}$ at $T = 5$~K and in magnetic fields $\mu_0 H$ applied along $\left[001\right]$, as indicated in the bottom left of each image.}\label{fig11_All}
\end{figure}

Fig.~\ref{fig11_All} shows a set of $B_z^{ac}(x,y)$ images as the sample is taken from one FP phase to the other across zero field. After cooling the sample in the absence of an applied field (zero-field cooling), we apply $\mu_0 H = 300$~mT to reach the FP phase. We then ramp the field down to $\mu_0 H = 136$~mT. Until this field, only the faint magnetic features shown in Fig.~\ref{fig11_All}a are visible. We attribute these fluctuations in $B_z^{ac}$ to the FP phase encountering surface roughness, because their small amplitude and lateral extent match features observed in atomic force microscopy images of the sample surface (see \ref{AppA1}). 

At 130~mT, a few ribbon-like features emerge, expanding as $H$ is further reduced. The measured stray-field patterns indicate that, within these domains, the magnetization is modulated with a period of about 130~nm propagating perpendicular to their long axis. Furthermore, as is visible at 124~mT, curling finger-like features are observed at the extrema of many of these domains. Reducing $H$, these elongated features expand, progressively populating the FP background until just below 106~mT, where the modulated phase subsumes the last FP regions. From $H=104$ to 74~mT, the modulations fill the entire scanning window and propagate nearly along $\left[110\right]$, suggesting a complete transition to a single-domain TS phase~\cite{qian_new_2018,chacon_observation_2018}. This phase should have a modulation period given by the wavelength of the intrinsic helimagnetic order in Cu$_2$OSeO$_3$, $\lambda = 62$~nm~\cite{adams_long-wavelength_2012}, and its propagation vector should point between $\left[111\right]$ and the applied field direction $\left[001\right]$, tilting towards the field as it is decreased~\cite{leonov_reorientation_2023}. The observed modulation, which propagates approximately along $\left[110\right]$ and whose period increases from 130 to 150~nm as a function of decreasing $H$, is consistent with the expected surface projection of the TS phase~\cite{milde_field-induced_2020}.


Around 70~mT, the stray field modulations indicative of the TS phase become unstable, changing over the time-scale of an individual image (tens of minutes), as in Fig.~\ref{fig11_All}h. Below this field, maps of $B_z^{ac}(x,y)$ stabilize and appear more complex than the single-frequency sinusoidal modulation observed at higher fields. Nevertheless, the dominant spatial period continues to increase as $H$ is reduced through zero and into reverse fields. As discussed in the next sections, these stray-field features appear to be a surface state related to the C and out-of-plane H phases expected in the bulk in this field range. They also resemble what was observed by Milde et al.\ via MFM at the surface of another Cu$_2$OSeO$_3$ sample in the C phase~\cite{milde_field-induced_2020}.

At -80 mT, the system again appears unstable on the time-scale of an individual image, indicating another nearby phase transition. In fact, at -85~mT, a modulation with a single spatial frequency abruptly reappears throughout the scanning area. Upon increasing the reverse field, the $B_z^{ac}(x,y)$ patterns mirror the behavior in forward applied field, with the modulation period shortening as a function of increasing reverse field up to -110~mT. Once again, this behavior is consistent with the surface projection of a TS phase, whose propagation vector tilts away from the applied field direction and towards $\left[111\right]$ as a function of increasing field. 

At -110~mT, thin regions of uniform contrast, consistent with the FP phase, begin to emerge within the modulated phase. As the TS phase breaks up and makes space for the FP phase, finger-like features reappear where the modulated domains fragment. At -125 mT, this fragmentation results in domains of FP, TS, and patterns consistent with disordered clusters of skyrmions within a FP background. The dipolar patterns in $B_z^{ac}(x,y)$ as well as the corresponding bright spots in $B_z(x,y)$ (see \ref{AppA2}) represent the reduction in $\lvert B_z\rvert$ caused by isolated configurations, whose core magnetization opposes the surrounding FP phase. The profiles are consistent with what is expected for disordered clusters of skyrmions, although we cannot distinguish whether they are N\'eel- or Bloch-type. TS domains gradually break up between -125 and -134 mT, leaving in their wake a LTS phase consisting of clusters of skyrmions in an FP background. Further application of reverse field results in a reduction in the density of the skyrmions. Finally, at -142~mT, the last few skyrmions are visible before being subsumed in a uniform FP background at higher reverse fields.

\subsection{Tilted spiral phase}\label{sec2}

\begin{figure}[t]%
\centering
\includegraphics[width=0.9\textwidth]{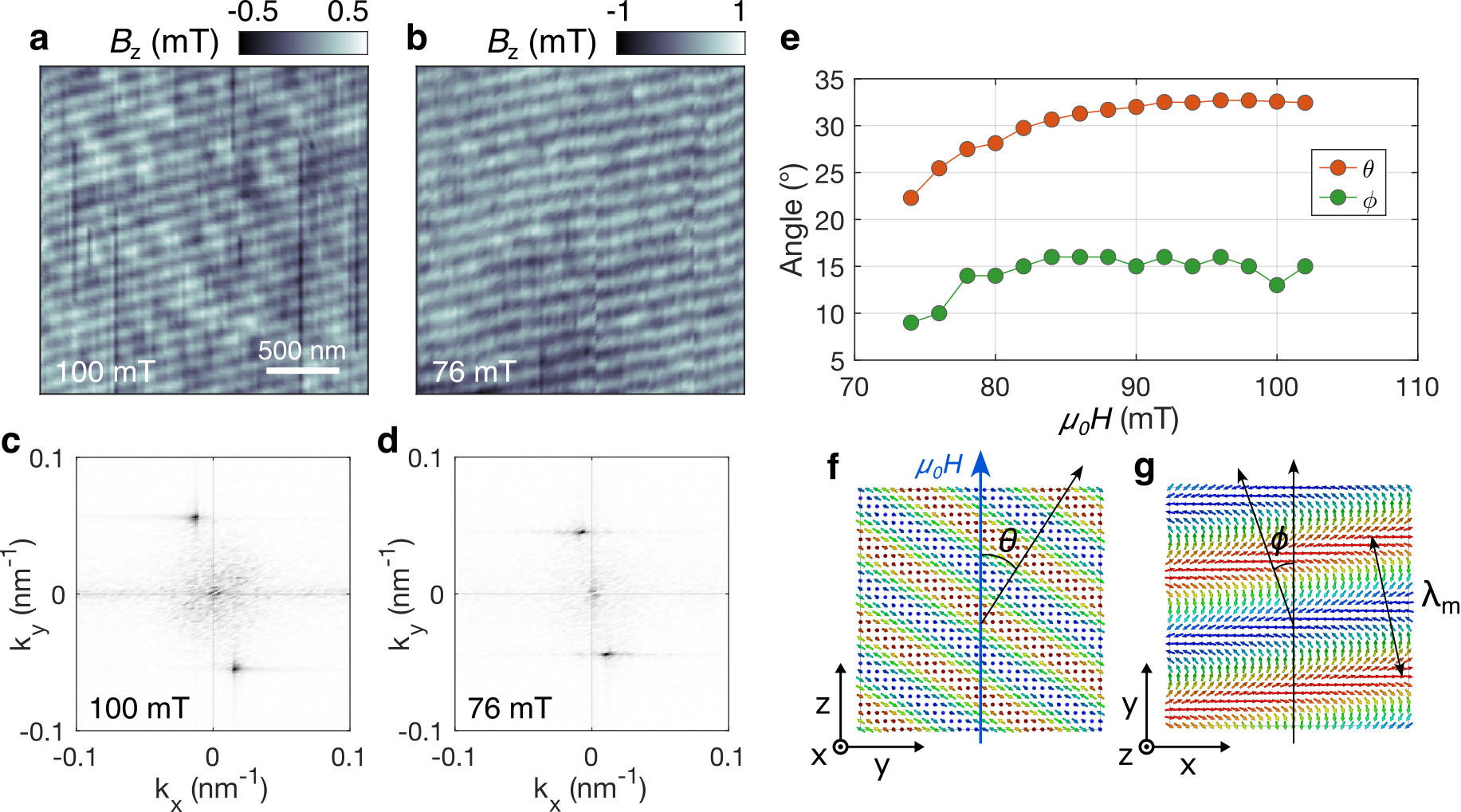}
\caption{Tilted spiral phase at the surface of Cu$_2$OSeO$_3$. \textbf{a},~Image of $B_z(x,y)$ at $T = 5$~K under indicated magnetic fields applied along $\left[001\right]$. \textbf{c} and \textbf{d},~Fast Fourier transforms of images a and b, respectively. \textbf{e},~Extracted tilt angles $\theta$ and $\phi$ as a function of the applied magnetic field. \textbf{f} and \textbf{g},~Illustration of the magnetization in the TS phase from the side and top of the crystal, respectively, with tilt angles $\theta$ and $\phi$, the wavelength of helimagnetic order $\lambda$, and the modulation wavelength $\lambda_m$, represented.}\label{fig22_TS}
\end{figure}

We next compare the modulated phase appearing between $\mu_0H = 104$ and 74~mT and again between -85 and -110~mT with the contrast expected from a TS phase intersecting the surface. For a TS phase propagating along the direction defined by the angle $\theta$ from the applied field direction and $\phi$ from $\hat{y}$, as shown in Figs.~\ref{fig22_TS}f and g, periodic magnetic features should appear on the surface with a modulation wavelength $\lambda_m = \lambda / \sin{\theta}$. By taking the two-dimensional fast Fourier transforms of $B_z(x,y)$, as shown in Figs.~\ref{fig22_TS}a-d, we extract $\theta$ and $\phi$ for each value of the applied field. As the field decreases from 104 to 74~mT, $\lambda_m$ lengthens from 108 to 137~nm, corresponding to $\theta$ tilting towards the field direction. The dependence $\theta(H)$, plotted in Fig.~\ref{fig22_TS}e, is similar to that of the TS phase measured by Chacon et al.\ via SANS in another Cu$_2$OSeO$_3$ sample~\cite{chacon_observation_2018}. At the same time, $\phi$ also changes, shifting from 15 to $10^\circ$ with respect to $\hat{y}$. This non-zero starting value for $\phi$ and its dependence on $H$ is likely due to the small misalignments of $\left[110\right]$ with $\hat{y}$ and of the applied field direction with $\hat{z}$.  As a result, the plane in which the TS propagation vector rotates, which is spanned by $\left[111\right]$ and the field direction, is slightly tilted with respect to the $xz$-plane. Overall, the behavior of the uniformly modulated phase both in forward and reverse applied field is consistent with what is expected from a projection of the TS phase on the surface.  

\subsection{Low-field phases and surface state}\label{sec3}

\begin{figure}[t]%
\centering
\includegraphics[width=1.0\textwidth]{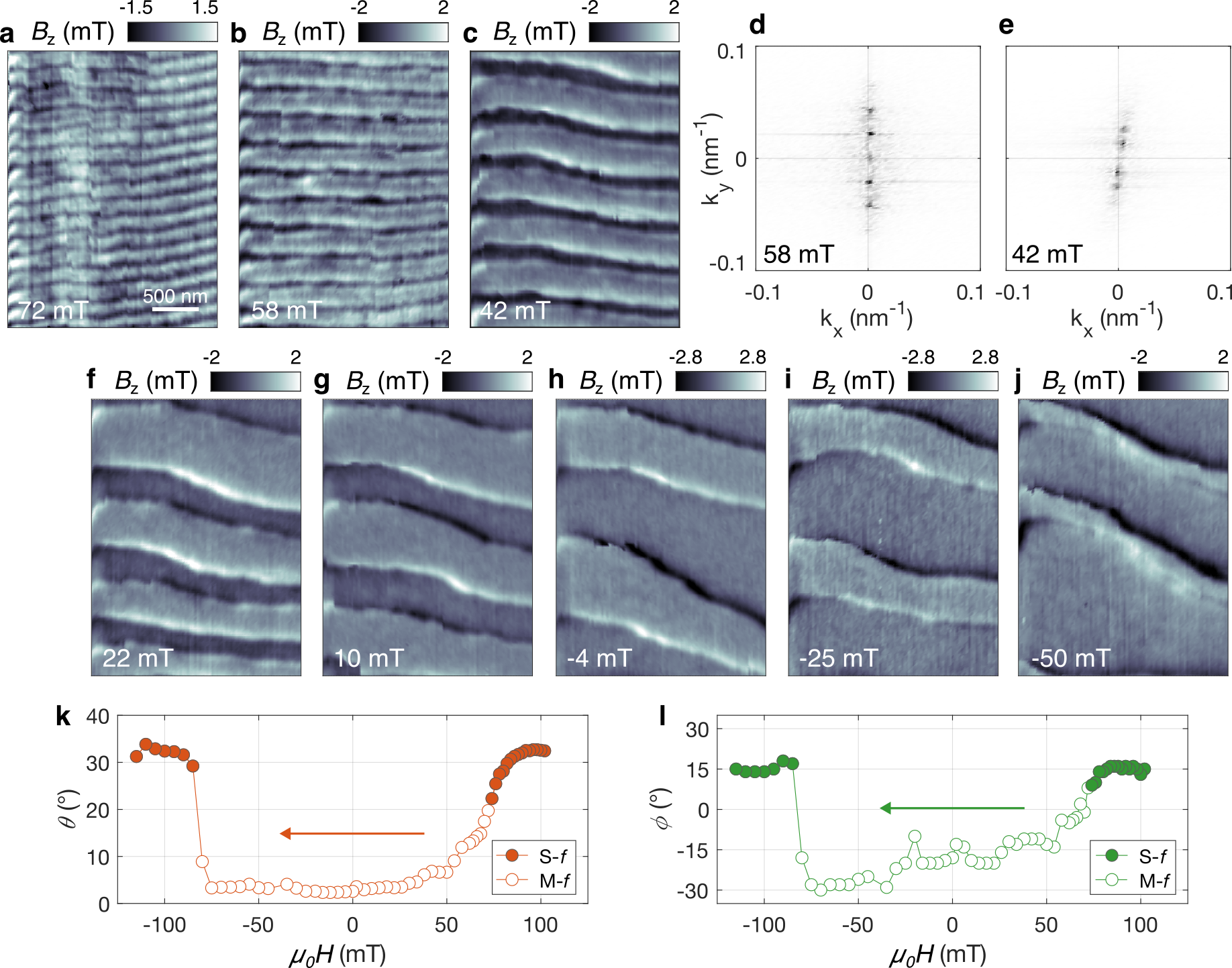}
\caption{Imaging magnetic modulation at the surface of Cu$_2$OSeO$_3$ in low applied field. \textbf{a},~Images of $B_z$ at $\mu_0 H = 72$~mT, \textbf{b},~58~mT and \textbf{c},~42~mT applied along $\left[001\right]$. \textbf{d-e},~FFT of images b and c, respectively. \textbf{f-j},~Images of $B_z$ at $\mu_0 H = 22$~mT, 10~mT, -4~mT, -25~mT, -50~mT. \textbf{k-l},~Extracted tilt angles $\theta$ and $\phi$ as a function of applied field. Filled and open points correspond to fields at which FFTs show a pattern with single-frequency (S-\textit{f}) modulation or multi-frequency (M-\textit{f}) modulation.}\label{fig33_SS}
\end{figure}

As the field is decreased below 72~mT, $B_z(x,y)$ maps begin to show a more complex periodic pattern, shown in Fig.~\ref{fig33_SS}. Fourier analysis reveals the appearance of a second spatial harmonic. Despite this additional complexity, which may be the signature of a stripe state stabilized at the surface, the dominant spatial period $\lambda_m$ continues to increase as the applied field is reduced to zero and into reverse field. Both $\theta(H)$ and $\phi(H)$, plotted in Fig.~\ref{fig33_SS}k and l, continue to reflect a gradual tilting of the magnetic order's propagation vector towards the applied field direction. This behavior is consistent with a TS phase, whose propagation vector is gradually aligning with an applied field that is slightly misaligned with the surface normal $\hat{z}$. Such a reorientation represents the transition from the TS to the C phase. Near zero field, $\theta$ settles around $2^\circ$, while $\phi$ approaches $30^\circ$ in reverse field. In this low-field regime, below about 40~mT, the C phase becomes an out-of-plane H phase, whose propagation direction is aligned along $\left[001\right]$. $\theta \simeq 2^\circ$ is consistent with the possible misalignment of the polished surface from $\left(001\right)$.  In reverse field, the discontinuities in $\theta(H)$ and $\phi(H)$ at $\mu_0 H = -80$~mT correspond to the sudden reappearance of the single-frequency TS phase with $\theta \simeq 30^\circ$ in reverse field. This asymmetry in applied field dependence reflects the strong hysteresis in the sample's magnetic behavior.

To characterize the microscopic magnetic states in the bulk, we carry out SANS measurements on the same sample studied by SSM using the same temperature and field protocol (see \ref{AppB}). The results are consistent with the interpretation of the SSM images: SANS patterns show scattering intensity corresponding to the TS phase below 100~mT, which gradually reduces its tilt angle and gives way to a C phase below 50~mT. This field corresponds to the region in the SSM measurement where $\theta < 5^\circ$, as shown Fig.~\ref{fig33_SS}k, and the propagation direction is nearly aligned to the applied field and $\left[001\right]$. As the field is further reduced, below 20~mT and down to -40 mT, SANS contrast points to the presence of a multi-domain H phase with propagation directions along the three $\left<100\right>$ directions. 

\begin{figure}[t]%
\centering
\includegraphics[width=0.85\textwidth]{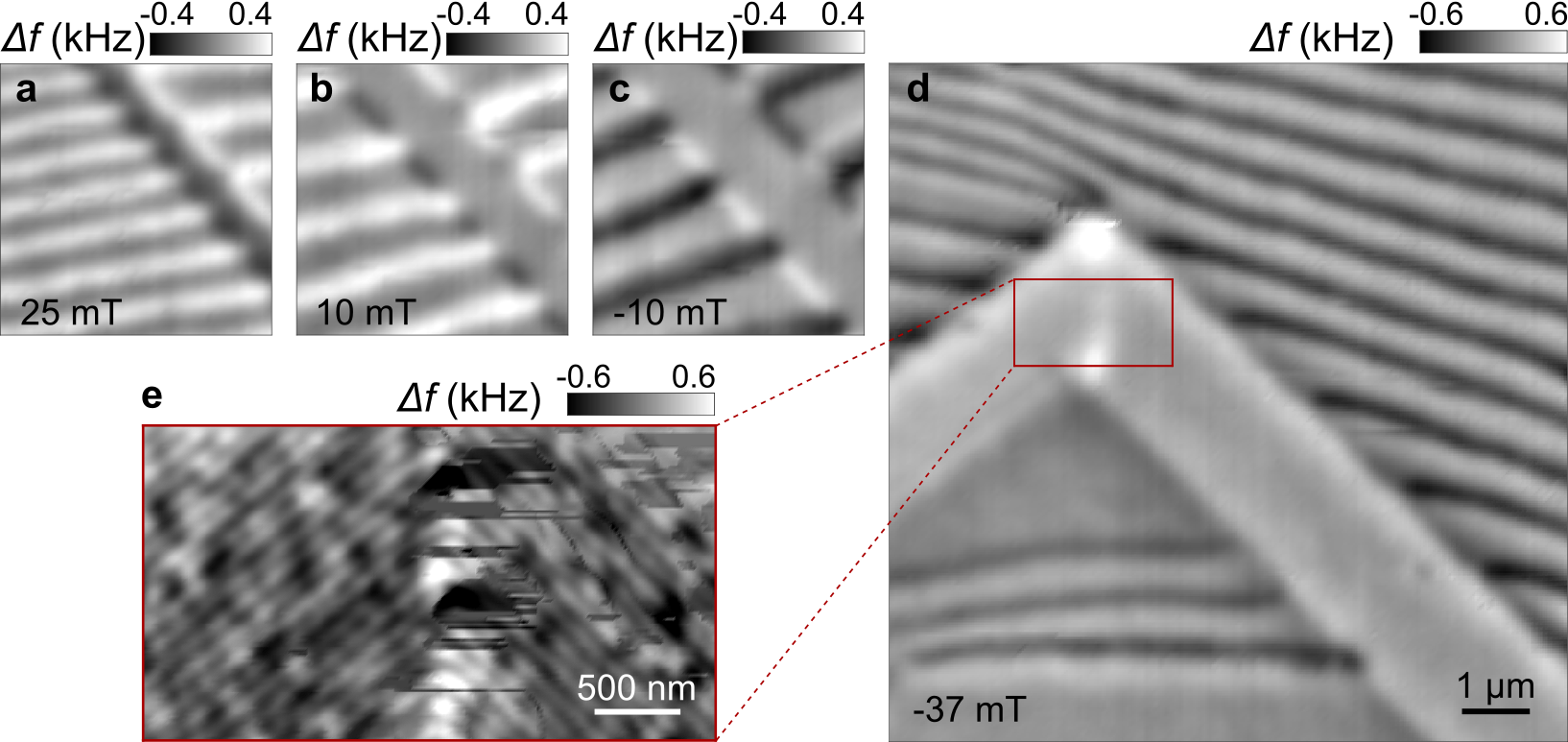}
\caption{MFM using a magnet-tipped NW at the surface of Cu$_2$OSeO$_3$. \textbf{a-d},~Images of $\Delta f$ at $T = 5$~K under indicated magnetic fields applied along $\left[001\right]$. \textbf{e},~Inset of d showing the helix modulation.}\label{fig44_NWMFM}
\end{figure}

Although no evidence of in-plane H phases is present in our SSM measurements, they do appear in additional high-resolution magnetic imaging by MFM using a magnet-tipped nanowire (NW) as the scanning probe~\cite{rossi_magnetic_2019,mattiat_nanowire_2020,mattiat_mapping_2024}. Such probes are ideal for imaging weak magnetic field patterns on the nanometer-scale~\cite{marchiori_nanoscale_2022}, because of their tiny magnetic tips, which are grown by focused-electron-beam-induced deposition of Co, and because of the NW's high-force sensitivity. Images of NW frequency shift $\Delta f (x,y)$, which is proportional to a combination of spatial derivatives of the sample's in-plane stray field $B_{x,y}(x,y)$ (see \ref{AppA3}), are shown in Fig.~\ref{fig44_NWMFM}. Below applied fields of 70~mT, $\Delta f (x,y)$ maps exhibit patterns of the same form and characteristic length scale as those observed via SSM. As discussed, these regions appear to be a surface manifestation of the C and out-of-plane H phases, whose propagation directions are slightly misaligned from the surface normal. Between 25~mT and -40~mT, however, $\Delta f (x,y)$ images also show the appearance and growth of new domains, as shown in Fig.~\ref{fig44_NWMFM}a-d. A high-resolution scan of these domains, shown in Fig.~\ref{fig44_NWMFM}e, reveals the presence of magnetic contrast consistent with in-plane H phases along $\left[100\right]$ and $\left[010\right]$ with $\lambda_m \approx \lambda = 62$~nm. The boundary between these two in-plane H domains appears to consist of curvature walls~\cite{schoenherr_topological_2018,leonov_topological_2022}. Therefore, we conclude that the images show the transition from a C to a multi-domain H phase, consisting of large domains oriented nearly out-of-plane and smaller domains of the two in-plane orientations. The absence of such multi-domain features in our SSM images is likely due to the small number and size of the areas that were investigated. 

\begin{figure}[t]%
\centering
\includegraphics[width=0.85\textwidth]{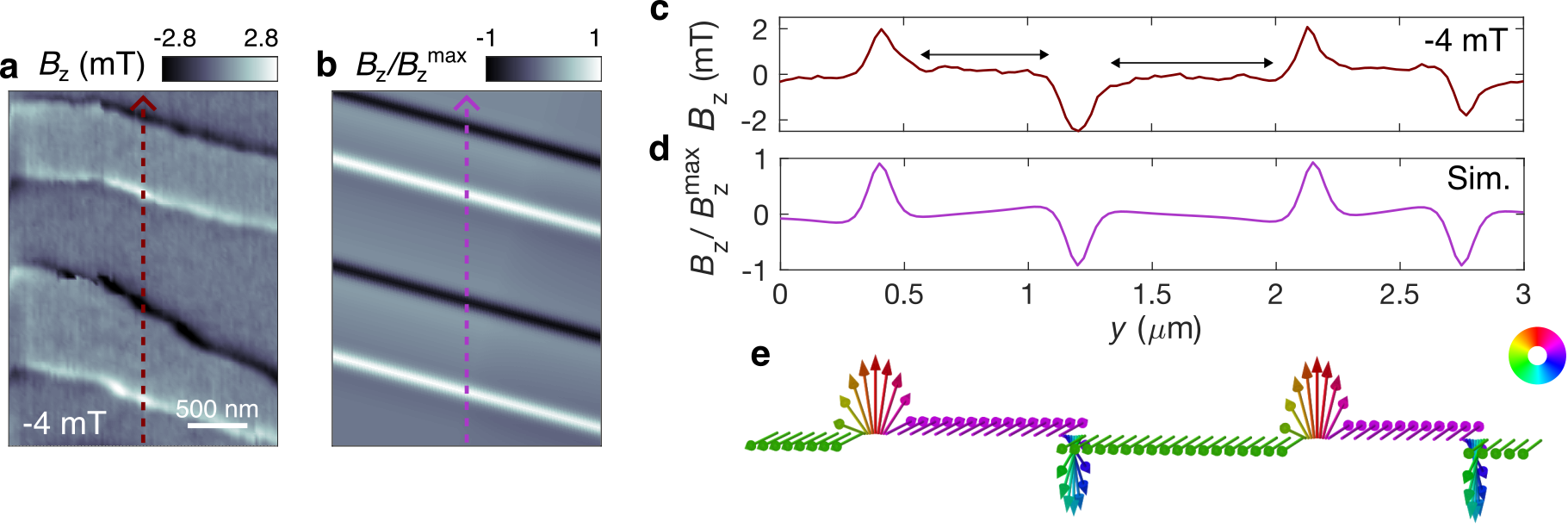}
\caption{Magnetic stripes at the surface of Cu$_2$OSeO$_3$. \textbf{a},~Measured $B_z(x,y)$ near zero field (at $\mu_0 H = -4$~mT) as also shown in Fig.~\ref{fig33_SS}g. \textbf{b},~The corresponding calculated stray field map assuming an idealized magnetic stripe state. \textbf{c-d},~Line-cuts taken along the dotted lines in \textbf{a-b}, respectively. \textbf{e},~Schematic diagram of an idealized magnetic stripe state.}\label{fig33_SS_B}
\end{figure}

The appearance of the stripe-like pattern characterized by a second spatial harmonic in the magnetic modulation below 72~mT, shown in Fig.~\ref{fig33_SS}, may be the result of the broken symmetry at the surface. Although Rybakov et al. postulated the presence of a stacked spin spiral phase at the surface of a chiral magnet coexisting with the C phase in the bulk, its wavelength and field dependence do not match our features~\cite{rybakov_new_2016}. The observed stray field modulation is most consistent with the in-plane magnetic stripe state described by Osorio et al.~\cite{osorio_chiral_2023}, especially near zero applied field. This stripe state is a distorted helical modulation consisting of long stretches of in-plane magnetization interrupted by short out-of-plane magnetization rotations, as illustrated in Fig.~\ref{fig33_SS_B}d. Fig.~\ref{fig33_SS_B} shows a comparison between the measured stray fields near zero field and simulated stray fields due to such an idealized stripe configuration, showing excellent agreement, especially compared to other distorted helical states, such as the chiral soliton lattice~\cite{okamuraEmergence2017,nakajima_uniaxial-stress_2018,osorio_chiral_2023}. In addition, we observe that as the field is swept from positive to negative, the pattern, which is initially characterized by wide bright stripes and narrow dark stripes, gradually transforms into a pattern of narrow bright stipes and wide dark stripes, as shown in Fig.~\ref{fig33_SS}f-j. This behavior, combined with an increase in the contrast between bright and dark, suggests that -- in an out-of-plane field -- the in-plane stripes develop a small out-of-plane component to their magnetization.  Nevertheless, for a complete understanding of this surface state, a detailed comparison to micromagnetic simulations should be carried out, which is beyond the scope of this study.

\subsection{Low-temperature skyrmion phase}\label{sec4}

\begin{figure}[t]%
\centering
\includegraphics[width=0.85\textwidth]{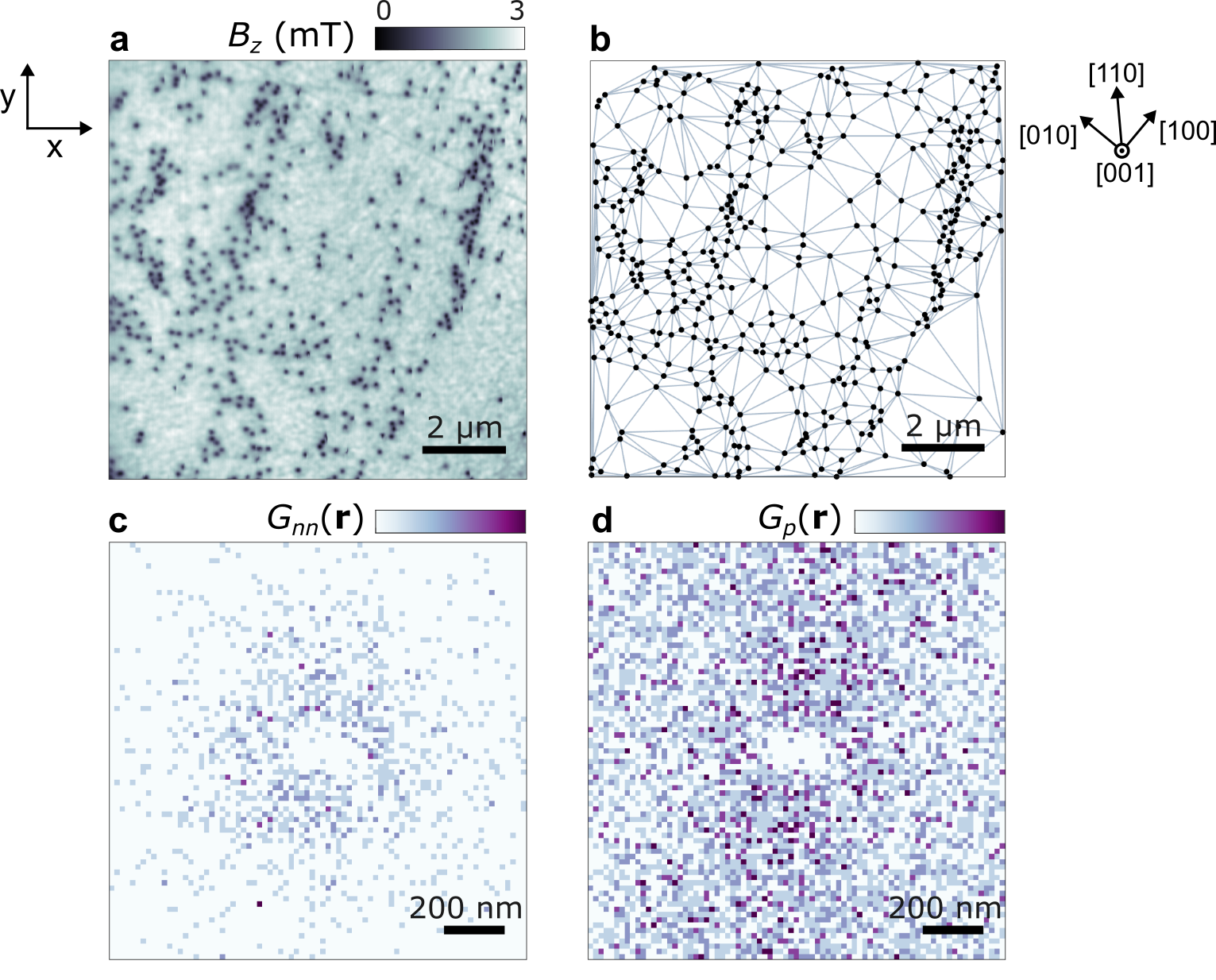}
\caption{Imaging magnetic skyrmions at the surface of Cu$_2$OSeO$_3$. \textbf{a},~Image of $B_z$ showing skyrmions. The faint patterns observed in the background are due to the roughness of the sample surface, which -- even in the FP state -- produces small fluctuations in $B_z$. \textbf{b},~Locations of individual skyrmions extracted from \textbf{a} with a triangular Delaunay mesh connecting the nearest neighbors. \textbf{c},~Nearest neighbor distribution function $G_{nn}(\mathbf{r})$ extracted from the triangulation edges. \textbf{d},~Skyrmion pair distribution function $G_{p}(\mathbf{r})$.}\label{fig2}
\end{figure}

The identification of the disordered clusters of skrymions in the magnetic images starting below -125~mT is consistent with an assosciated SANS contrast measured in the same sample (see \ref{AppB}). Fig.~\ref{fig11_All}p shows that skyrmion clusters emerge in a FP background from domains of the TS phase, which break up in an increasing reverse applied field, along the lines of what was suggested by Leonov and Pappas~\cite{leonov_topological_2022}. The tendency of skyrmion clusters to form along the $\left[110\right]$ propagation direction of the TS phase, evident in Fig.~\ref{fig2}a, is likely a vestige of this process.

A map of the skyrmion positions extracted from the image in Fig.~\ref{fig2}a along with the Delaunay triangulation connecting these positions, is shown in Fig.~\ref{fig2}b. Using the distances between skyrmions determined by the edge of the triangular cells, we determine the distribution of nearest neighbor distances, $G_{nn}(\mathbf{r})$, shown in Fig.~\ref{fig2}c. The distribution shows a broad ring-like feature indicating an average nearest-neighbor distance between 100 and 400~nm. Based on the extracted skyrmion locations, we also construct a pair distribution function, $G_{p}(\mathbf{r})$, showing the probability of finding a skyrmion a distance $\mathbf{r}$ from another skyrmion. $G_{p}(\mathbf{r})$ is plotted in Fig.~\ref{fig2}d and shows a distinct absence of both translational and orientational order (see \ref{AppC}). The lack of contrast for $\lvert \mathbf{r} \rvert \lesssim 100$~nm is mostly the result of the $\sim$100-nm spatial resolution of our SSM probe. Its asymmetry reflects the formation of clusters roughly aligned along the $y$-direction. 
The presence of disordered skyrmion clusters and the absence of a lattice within a FP background is consistent with expectation~\cite{bannenberg_multiple_2019,baral_direct_2023}: as the system exits the TS state and enters the FP state, skyrmions remain stable, but the skyrmion lattice does not form because of skyrmion-skyrmion repulsion. SANS measurements of this sample also show a weak circular signature, indicating an LTS phase with a strong disorder. 

\subsection{Skyrmion manipulation}\label{sec5}

\begin{figure}[t]%
\centering
\includegraphics[width=0.95\textwidth]{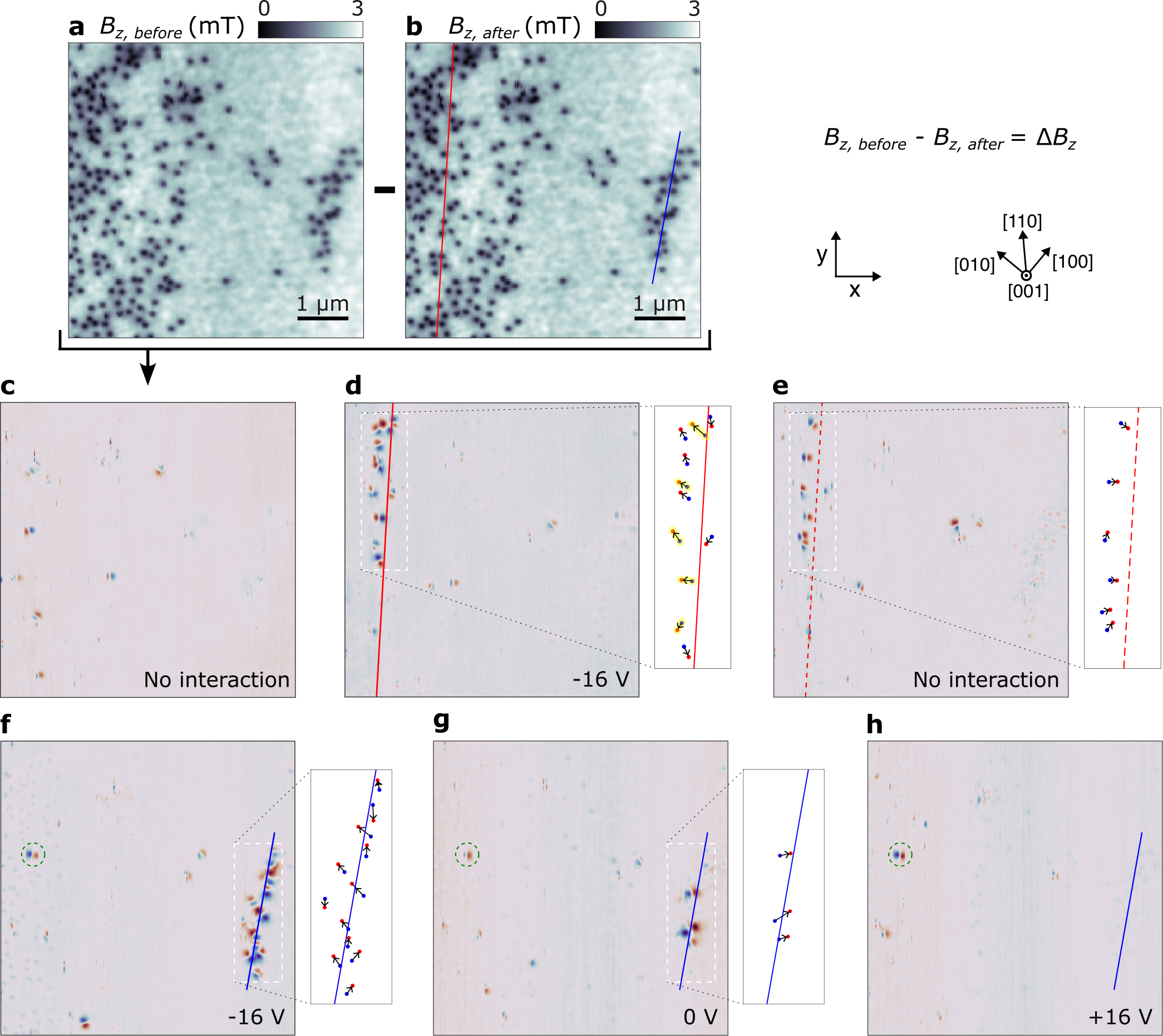}
\caption{Manipulating skyrmions with a charged tip. \textbf{a-b},~Images of $B_z$ taken sequentially showing a disordered array of skyrmions before and after. Red and blue lines indicate paths along which the probe is scanned in close proximity to the sample with $V_{\text{tip}}$ applied. \textbf{c-h},~Images of $\Delta B_z$ showing the difference in the stray field pattern between two sequential images interrupted by: \textbf{c},~no probe interaction between images, \textbf{d},~the probe scanned along the red line with $V_{\text{tip}} = -16$~V, \textbf{e},~no probe interaction after the latter interaction, \textbf{f},~the probe scanned along the blue line with $V_{\text{tip}} = -16$~V, \textbf{g},~the probe scanned along the blue line with $V_{\text{tip}} = 0$~V, and \textbf{h},~the probe scanned along the blue line with $V_{\text{tip}} = +16$~V. Insets show the change in the position of individual skyrmions extracted from the images. Blue points correspond to the original skyrmion position, and red to the new position.}\label{fig3}
\end{figure}

In order to determine the stability of the observed skyrmion configurations, images of the same region, which is shown in Fig.~\ref{fig3}a-b, are taken in succession. The difference $\Delta B_z(x,y)$ of two successive images, shown in Fig.~\ref{fig3}c, demonstrates the stability of the entire configuration: under 10 near-surface skyrmions change position out of over 444 in the image and their average displacement is $106\pm23$~nm. These skyrmions are likely perturbed from shallow pinning sites by thermal fluctuations and subsequently fall into other nearby sites~\cite{nozaki_brownian_2019,zazvorka_thermal_2019,zhao_topology-dependent_2020}. A few such rearrangements are caught during the imaging process itself, in Figs.~\ref{fig2}a, \ref{fig3}a, and b, resulting in the occasional skyrmion appearing as a circular spot with a discontinuous cut along the $y$-axis (fast scanning axis). Unlike MFM, whose magnetic tip produces a stray field at the sample surface large enough to affect the local magnetic phase~\cite{milde_field-induced_2020}, the SQUID-on-tip probe produces negligible stray fields. Any interaction between the scanning probe and the skyrmions is, therefore, the result of differences in electric potential between the probe and the sample. The resultant electric field may then act on the skyrmions via the sample's magnetoelectric coupling. To avoid any such interactions during imaging, we raster at a large enough tip-sample distance (80-100~nm) such that no probe-induced effects are observed. The highest stability configurations, such as the one studied here, are found at larger $H$, closer to the transition to the FP state. At lower fields, closer to the nucleation of the skyrmions and their coexistence with both a FP background and TS domains, skyrmions are observed to be less stable and more frequently hop between pinning sites. 

Next, we exploit the magnetoelectric coupling in Cu$_2$OSeO$_3$ to manipulate individual skyrmions. By applying a voltage $V_{\text{tip}}$ to the SQUID-on-tip with respect to the sample, we induce a localized electric field beneath the tip and use it to address individual skyrmions. Fig.~\ref{fig3}d shows $\Delta B_z$ of images taken before and after scanning the SQUID-on-tip along the line shown in red at a tip-sample distance of 40~nm and a rate of 50~nm/s with $V_{\text{tip}} = -16$~V.  11 dipolar patterns appearing near the top part of the tip's path -- highlighted in the corresponding inset -- indicate the displacement of nearby skyrmions, by an average of $120\pm30$~nm. The clustering of these displacements near the top part of the tip's path is the result of a small tilt between the sample and the tip path, resulting in a smaller tip-sample spacing by approximately 10~nm at the top compared to the bottom of the path. This smaller displacement, in turn, results in a larger localized electric field at the sample surface. Away from the interaction path, 6 skyrmions also rearrange, presumably due to thermal fluctuations. Fig.~\ref{fig3}e shows $\Delta B_z(x,y)$ of two images consecutively taken, after the interaction with the charged tip and without any further perturbations. This image, which is a measure of the stability of the configuration after interaction with the tip, shows 6 dipolar signals near the original region of interaction with an average displacement of $101\pm11$~nm, along with a few randomly distributed thermal rearrangements away from this area. The signals near the original interaction line occur at the same positions, highlighted in yellow in the inset of Fig.~\ref{fig3}d, and have the opposite polarity as the displacements induced by the charged tip. These displacements represent skyrmions relaxing back to positions close to the ones before the perturbation. Having been displaced from deeper to shallower pinning sites, they relax back to the original or nearby sites under the influence of thermal fluctuations. The displacement of another cluster of skyrmions is shown in Fig.~\ref{fig3}f after a second scan of the SQUID-on-tip along a different path with the same applied voltage and at a slightly closer tip-sample distance of 30~nm. We observe a larger average skyrmion displacement of $133\pm32$~nm with 12 dipolar signals shown in the inset. Repeating the experiment with the same tip-sample distance and $V_{\text{tip}} = 0$~V results in fewer displaced skyrmions, as shown in Fig.~\ref{fig3}g. Positive applied tip voltages, rather than leading to symmetric skyrmion displacements, produce no displacements, as shown in Fig.~\ref{fig3}h, which is taken after scanning the tip with $V_{\text{tip}} = +16$~V. In addition, away from the interaction region, one dipolar signal is persistently seen in Fig.~\ref{fig3}f-h, which is indicated by a dashed green circle, suggesting a fluctuating skyrmion located in a shallow pinning potential. 

In all maps of $\Delta B_z(x,y)$, skyrmions are displaced by an average of $116\pm27$~nm either by thermal fluctuations or by interaction with the charged tip. Skyrmions displaced by the tip do not appear to move along any preferential direction with respect to the crystal or the motion of the charged tip.  Also, all signals observed in $\Delta B_z(x,y)$ have a dipolar character, indicating that both thermal fluctuations and tip-induced interaction result in skyrmion displacement rather than their creation or annihilation. Simulations have suggested that skyrmions can be created in Cu$_2$OSeO$_3$ by a tip-induced local electric field, although the required electric fields are two orders of magnitude larger than those applied here~\cite{mochizuki_writing_2015}.

\section{Discussion}\label{sec6}

The overall picture given by our SSM and MFM images of the stray magnetic field patterns at the surface of Cu$_2$OSeO$_3$ is consistent with the projection of the magnetic phases known to be present in the bulk. As a function of decreasing field applied along $\left[001\right]$, we observe FP, TS, and a stripe pattern at the surface likely related to out-of-plane C and H phases slightly misaligned with the surface normal. At low fields, we image a multi-domain H phase and in large reverse fields -- at the boundary between TS and FP phases -- we observe the nucleation of disordered clusters of skyrmions. Furthermore, the tilt angle of the TS phase presents the expected dependence on applied magnetic field. 

The images do, however, reveal a number of important insights. The observation of the coexistence of domains of FP, TS, and disordered LTS phases near the transition between FP and TS phases is particularly noteworthy. The images of this process confirm the first-order nature of this transition for a field applied along $\left<100\right>$~\cite{leonov_reorientation_2023}. They also show that the rupture of TS domains in the FP background nucleates clusters of low-temperature skyrmions. Leonov and Pappas proposed two scenarios for skyrmion nucleation: (i) via rupture of in-plane H domains and (ii) from domain boundaries between different TS states~\cite{leonov_topological_2022}. We do not see evidence for either of these processes at the surface. In fact, we always observe a single-domain TS phase -- not domains of TS tilting towards different $\left<111\right>$-axes -- perhaps because one of the four directions is favored by the slight misalignment of the applied field with $\left[001\right]$. Nevertheless, the formation of stable skyrmion clusters at the boundaries between TS and FP domains does follow a scenario similar to those described by Leonov and Pappas. Certainly, the images provide a microscopic explanation for the hysteretic behavior of the LTS phase~\cite{leonov_field_2020,leonov_topological_2022}.


The effect of the surface appears in both SSM and MFM maps, which show evidence of a low-field stripe state. With the system either in the C or out-of-plane H phase, the stray magnetic field at the surface displays a complex modulation, whose fundamental spatial frequency is set by the propagation direction's slight misalignment from the surface normal. Signatures of this state are absent from SANS or other bulk measurements of Cu$_2$OSeO$_3$. Furthermore, the state's stray field pattern is consistent with an idealized in-plane stripe state.  Such a distorted helical state could be stabilized at the surface due to symmetry breaking at this boundary and its effect on the balance of magnetic energies. Effects related to surface strain~\cite{okamuraEmergence2017,nakajima_uniaxial-stress_2018} and surface defects may also play a role.

The experiments also highlight the importance of random pinning potentials in fixing the configuration of near-surface skyrmions. A number of mechanisms could be responsible, including structural defects or impurities in the bulk, surface defects or adatoms adhering to the surface, or local variations of either the DMI or magnetic anisotropy. Whatever the mechanism, the average density of pinning sites can be estimated from the average observed displacement of skyrmions resulting either from thermal or tip-induced perturbations, yielding 74~$\mu$m$^{-2}$. This density represents almost one quarter of the density corresponding to the bulk skyrmion lattice. 

A final question revolves around the mechanism by which the charged scanning probe displaces skyrmions, which relies on the coupling between the locally applied electric field and the emergent electric polarization in Cu$_2$OSeO$_3$. This coupling allows both for the application of a force on individual skyrmions~\cite{seki_magnetoelectric_2012} and for the local tuning of the relative energies of skyrmions compared to competing magnetic phases~\cite{huang_situ_2018}. In the former case, given that skyrmions in Cu$_2$OSeO$_3$ for $\mathbf{H} \parallel [001]$ are associated with an electric quadrupolar moment~\cite{seki_magnetoelectric_2012}, they are susceptible to forces induced by inhomogeneous electric fields, such as those created by a local probe. However, for this mechanism to be responsible for the observed displacements, the effect would have to be symmetric as a function of applied electric field. Our observations contradict this prediction: multiple skyrmion displacements are observed for negative values of tip voltage, e.g.\ $V_{\text{tip}} = -16$~V in Fig.~\ref{fig3}c and e, while none are observed for positive values, e.g.\ $V_{\text{tip}} = +16$~V in Fig.~\ref{fig3}g. Given the size of the observed asymmetry, which is apparent up to at least $\pm16$~V, differences in tip-sample contact potential are not sufficient to account for it.

On the other hand, experiments in bulk and lamellae of Cu$_2$OSeO$_3$ have shown that an applied electric field can tune the free energy of the skyrmion phase, thus either enhancing or suppressing its stability~\cite{kruchkov_direct_2018} depending on the sign of the field. In this way, starting from an initial helical phase, skyrmions were created via the application of an electric field~\cite{huang_situ_2018}. A similar mechanism could be at work in our experiment, where the local application of a $V_{\text{tip}} < 0$ could result in the suppression of a skyrmion's stability beneath the tip, perturbing it from its pinning potential and displacing it to a more favorable nearby position. The application of $V_{\text{tip}} > 0$ could then result in an enhancement of a skyrmion's stability beneath the tip, preserving its original position. It is interesting to note that throughout our experiments, no skyrmion creation or annihilation events were observed, despite local electric fields applied to the surface in excess of $10^7$~V/m. This observation provides experimental evidence for the robustness of magnetic skyrmions. The fact that the skyrmions can both be controllably displaced without being destroyed or created is promising for storage and information processing applications.  

\section{Conclusion}\label{sec7}

The presented SSM and MFM images give a clear picture not only of the low-temperature magnetic phases at the surface of Cu$_2$OSeO$_3$, but also of their phase transitions in a field applied along the easy axis. They shed light on the microscopic nucleation of low-temperature skyrmions from rupturing TS domains in the FP background, during the first-order transition from the TS to FP phase. We also find evidence for a magnetic stripe phase at the surface. Using a charged scanning probe to displace and image individual skyrmions in disordered clusters, we observe behavior dominated by pinning potentials. Together, these findings point to the strong effects of the surface in modifying the magnetic phases present in the bulk. The extent to which these effects are determined by intrinsic properties of the surface or imperfections due to defects, impurities, or roughness is a topic for future study.

Along with the insights about phases and phase transitions, our work also sheds light on the behavior of individual skyrmions at the surface of Cu$_2$OSeO$_3$ and other similar materials. In general, skyrmions occur in lattices or as isolated particles within a different magnetic phase~\cite{loudon_direct_2018}. Understanding how isolated skyrmions behave and how to manipulate them is crucial for potential applications. In particular, identifying their pinning mechanisms and how to engineer pinning sites is an important ingredient for any information storage or processing scheme. Although the demonstrated skyrmion displacements are small, experiments on thinner samples or with patterned surface electrodes could allow for the application of electric fields orders of magnitude larger than those applied here, potentially allowing for greater control. The possibilities are intriguing: for example, one may imagine patterning nanometer-scale gates at the surface of Cu$_2$OSeO$_3$ in order to control their position. Such an architecture is particularly attractive given the capabilities of nanometer-scale patterning technology and the fact that control via electric field can be carried out with negligible dissipation, unlike control via electric current or magnetic field.

\vspace{15mm}

\bmhead{Acknowledgments}
We thank Flaviano Jos\'e Dos Santos, Dirk Grundler, and Andrey Leonov for insightful discussions. We are grateful to Hinrich Mattiat for work setting up the nanowire MFM, Lorenzo Ceccarelli for help with the figures, and Sascha Martin and the machine shop of the Department of Physics at the University of Basel for their role in building the scanning probe microscopes. We acknowledge the support of the Canton Aargau; the Swiss National Science Foundation via Project Grant No. 200020-159893 and Sinergia Grant ``Nanoskyrmionics" (Grant No. CRSII5-171003); the University of Basel via its Research Fund (Project No. 4637580); and the Swiss Nanoscience Institute via its Ph.D. Grant P1905. This work is based partly on experiments performed at the Swiss spallation neutron source SINQ, Paul Scherrer Institute, Villigen, Switzerland.

\pagebreak

\appendix

\section{Appendix}

\subsection{Cu$_2$OSeO$_3$ sample}\label{AppA1}
\begin{figure}[H]%
\centering
\includegraphics[width=0.7\textwidth]{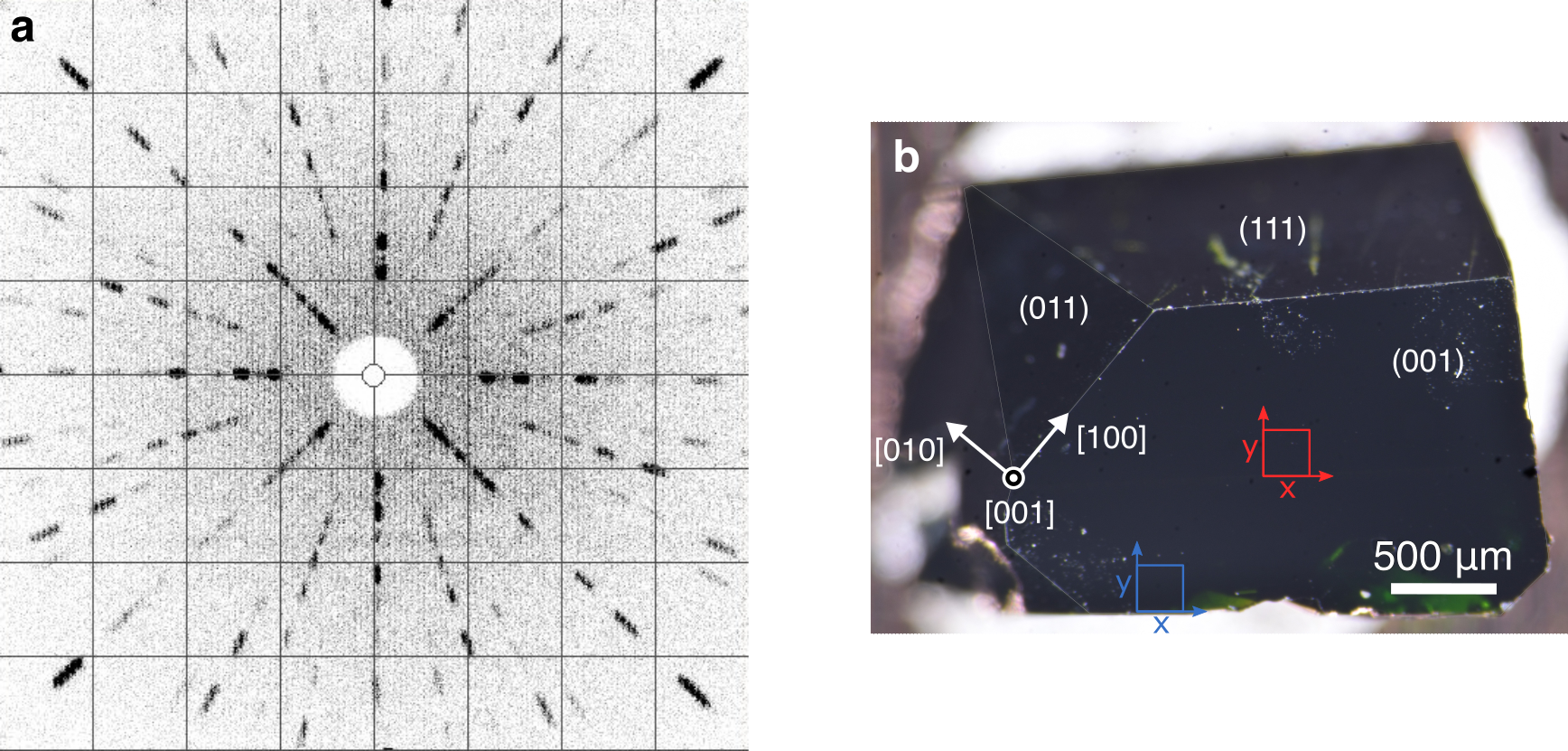}
\caption{Cu$_2$OSeO$_3$ crystallographic orientations. \textbf{a},~Four-fold symmetric diffraction pattern obtained using X-ray Laue on the scanned $(001)$ surface of the Cu$_2$OSeO$_3$ single crystal after mechanical polishing. \textbf{b},~Optical micrograph of the crystal with labeled crystallographic axes and planes. The location of the SSM scanning window is represented in red, and the MFM scanning window is in blue, with both sizes exaggerated for clarity. The edges of the crystal have also been highlighted for clarity.}\label{figS1_laue}
\end{figure}

\begin{figure}[H]%
\centering
\includegraphics[width=0.35\textwidth]{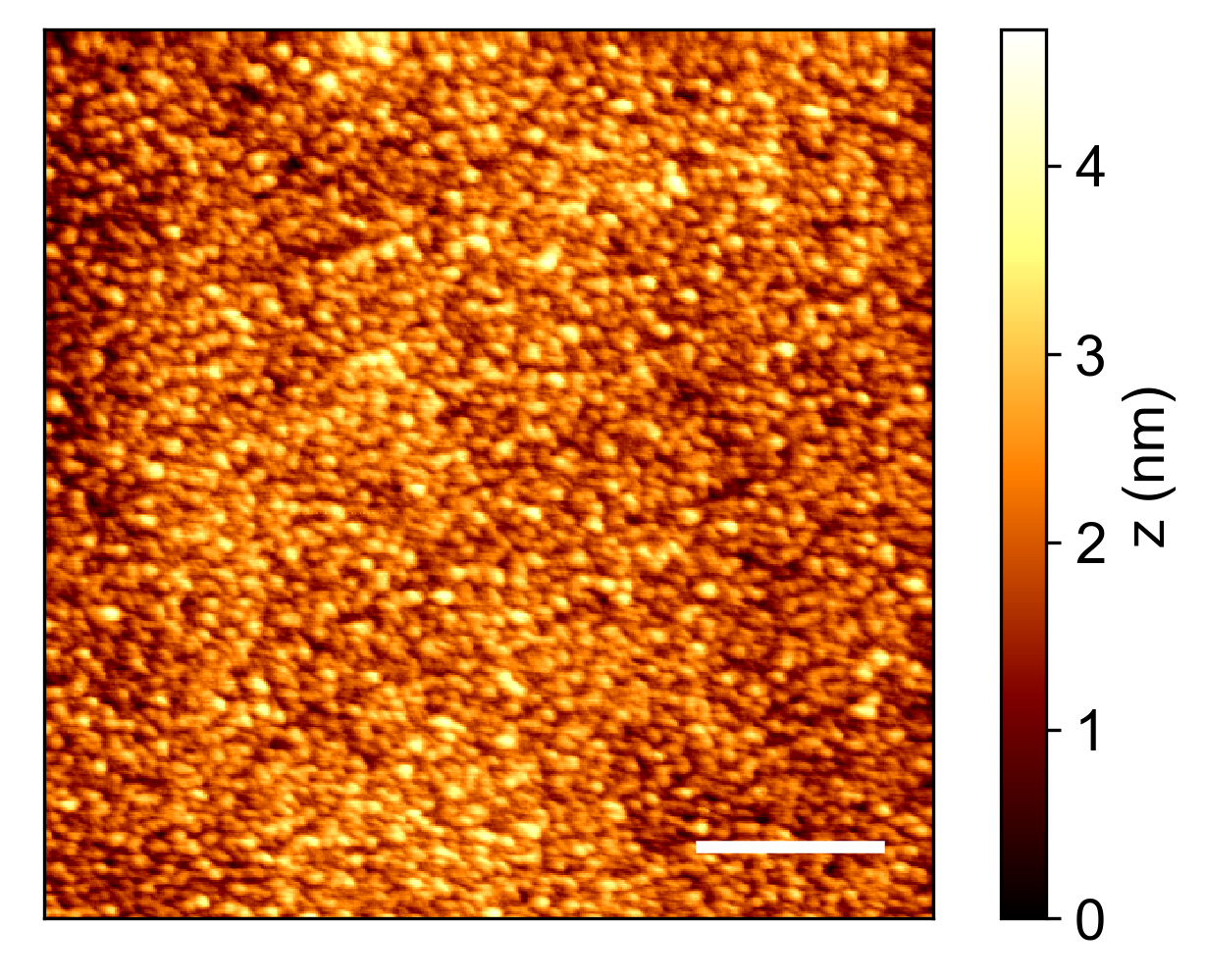}
\caption{Atomic force micrograph of the $(001)$ surface of the same Cu$_2$OSeO$_3$ crystal measured and described in the main text. The image shows an approximate roughness of 5~nm on a characteristic length scale of 70-130~nm.  Scalebar: 1~$\mu$m.}\label{figS2_afm}
\end{figure}

The Cu$_2$OSeO$_3$ crystals are grown via chemical vapor transport with CuO and SeO$_2$ as starting materials and HCl as the transport agent according to the process described by Baral et al~\cite{baral_tuning_2022}. A roughly rectangular single crystal  with dimensions $2.4 \times 1.9 \times 1.1$~mm$^3$ is chosen with a natural $\{100\}$ facet. The orientation is verified using a Laue camera. The facet is then mechanically polished and the orientation is subsequently re-verified using the Laue camera. As shown in Fig.~\ref{figS1_laue}a, the pattern is consistent with a $\{100\}$ surface. Fig.~\ref{figS1_laue}b shows the crystallographic directions and faces, which are identified based on the X-ray diffraction and measurements of the physical angles between natural faces.  The miscut angle is measured using rocking curves with XRD to be less than 1 degree. Before imaging by scanning SQUID microscopy (SSM), the $(001)$ surface is etched using Ar plasma, and the resulting roughness is measured by atomic force microscopy (AFM). As shown in Fig.\ref{figS2_afm}, the surface has a roughness of  5~nm with a lateral feature size of 70-130~nm. SSM, SANS, and MFM measurements presented here are carried out on the same Cu$_2$OSeO$_3$ crystal. Follow-up SSM measurements on another crystal confirm the reported behavior.

\subsection{Scanning SQUID microscopy}\label{AppA2}

\begin{figure}[b]%
\centering
\includegraphics[width=1.0\textwidth]{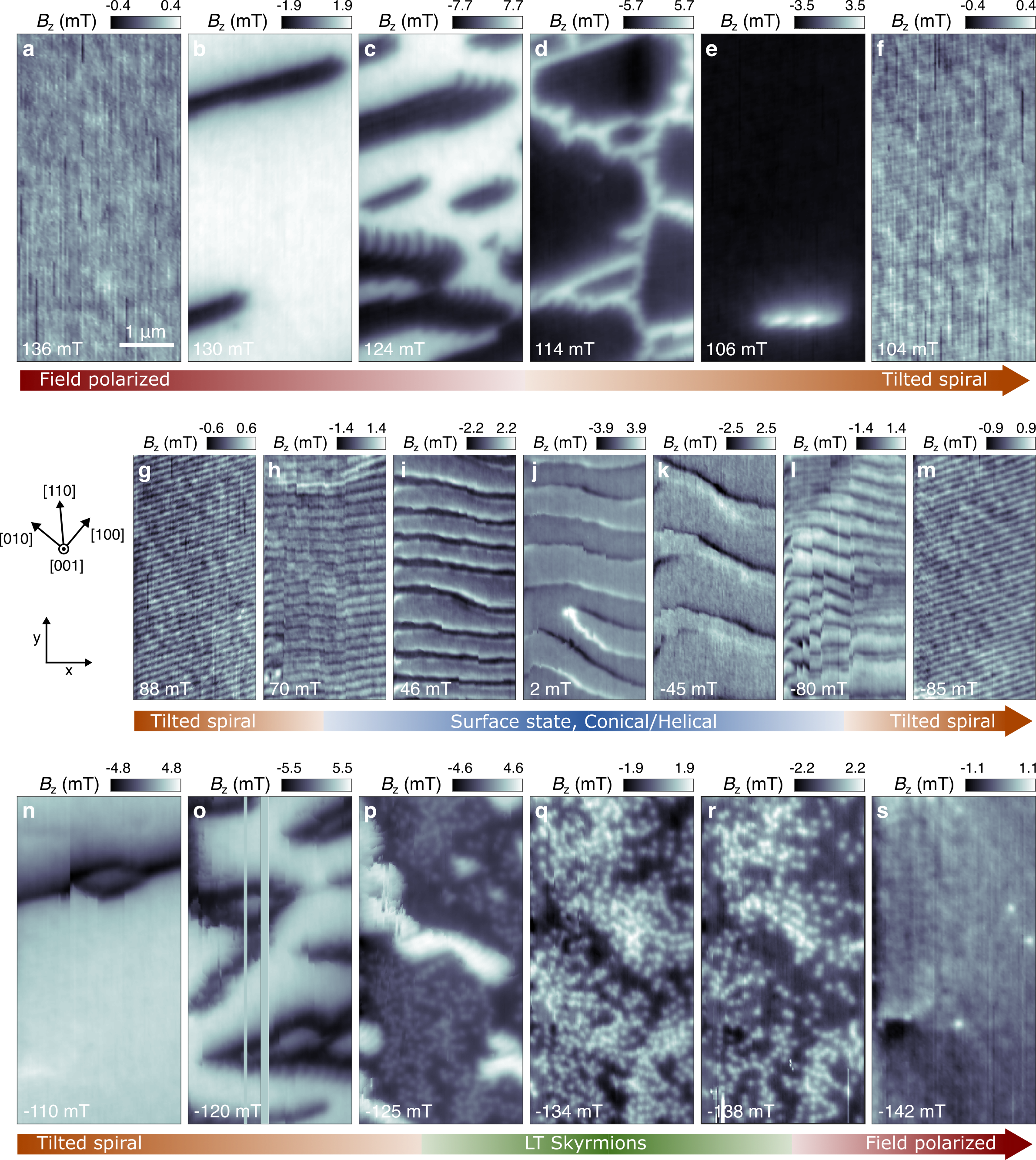}
\caption{SSM of magnetic phases at the surface of Cu$_2$OSeO$_3$. \textbf{a}-\textbf{s}~Images of $B_z$ at $T = 5$~K and in magnetic fields $\mu_0 H$ applied along $\left[001\right]$, as indicated in the bottom left of each image.}\label{figS5_Bzimg}
\end{figure}

We map the stray magnetic field perpendicular to the $(001)$ surface of a Cu$_2$OSeO$_3$ crystal using a SQUID-on-tip probe. The SQUID-on-tip is fabricated by sputtering either MoGe or Nb on the apex of a pulled quartz capillary as described by Romagnoli et al.~\cite{romagnoli_fabrication_2023}. The effective loop diameter of 80~nm and 120~nm of the SQUID-on-tip probes are extracted from measurements of the critical current $I_{\text{SOT}}$ as a function of a uniform magnetic field $H$, applied perpendicular to the loop. A serial SQUID array amplifier (Magnicon) is used to measure the current flowing through the SQUID-on-tip~\cite{romagnoli_fabrication_2023}.  

The crystal is mounted in a high-vacuum microscope operating at 5~K, with the $(001)$ surface just below the SQUID-on-tip probe and parallel to its scanning plane ($xy$-plane). Since the current response of the SQUID-on-tip is proportional to the magnetic flux threading through its SQUID loop, it provides a measure of the component of the local magnetic field perpendicular to $(001)$ surface of the sample $B_z$, integrated over the loop. By scanning the sample with piezoelectric actuators (Attocube) at a constant tip-sample spacing between 80 and 100~nm, we map $B_z$. The spatial resolution of the SSM is limited by this spacing and by the SQUID-on-tip diameter. 

The tip-sample spacing is controlled using a tuning fork mechanical resonator (Qplus), which is physically coupled to the body of the SQUID-on-tip at approximately 100~$\mu$m from the tip apex. A piezoelectric actuator and a phase-locked loop are used to excite and monitor the tuning fork’s mechanical resonance and tip-sample spacing. We drive the probe with the piezoelectric actuator on resonance, such that the SQUID-on-tip oscillates along the $y$-direction with a few nanometers of oscillation amplitude. By measuring the SQUID’s response at this frequency using a lock-in amplifier (Zurich Instruments MFLI), we also map $B_z^{ac}$, which is proportional to $dB_z/dy$. Maps of $B_z^{ac}$ show a greater level of detail than $B_z$, because of the reduction of noise offered by the lockin’s spectral filtering and the higher spatial resolution characteristic of measuring a magnetic field derivative~\cite{marchiori_nanoscale_2022}. On the other hand, $B_z$ shows whether the underlying magnetization points along or against the applied field (bright points along $\hat{z}$ and dark points along $-\hat{z}$) than $B_z^{ac}$. Fig.~\ref{figS5_Bzimg} shows maps of $B_z$ corresponding to the maps of $B_z^{ac}$ shown in Fig.~\ref{fig11_All}. $B_z$ and $B_z^{ac}$ are measured simultaneously using a scanning probe microscopy controller (Specs) at a scan rate of 0.2~$\mu\text{m}/\text{s}$, time per pixel of 100~ms.

\subsection{Nanowire magnetic force microscopy}\label{AppA3}

The MFM probes are high aspect ratio NWs, which are a few micrometers long and a few tens of nanometers in diameter. Their nanometer-scale Co magnetic tip is grown by focused-electron-beam-induced deposition~\cite{mattiat_mapping_2024}. We map shifts in the NW's mechanical resonance frequencies, which are produced by the tip–sample interaction while scanning above the sample surface at a constant height of $\sim$50~nm. During a scan, we record the frequency shift of both of the NW’s flexural modes, $\Delta f_1 = f_1 - f_{0,1}$ and $\Delta f_2 = f_2 - f_{0,2}$. $f_{0,1}$ and $f_{0,2}$ are the natural resonance frequencies of the modes in the absence of interaction with the sample. The quantity displayed in the images is the sum of these frequency shifts $\Delta f = \Delta f_1 + \Delta f_2$. In the limit of a strongly magnetized sample, each mode's frequency shift is proportional to the spatial derivative of the sample’s in-plane stray magnetic field taken along each of the NW mode directions: $\Delta f_{1,2} \propto dB_{x1,2}/dx_{1,2}$~\cite{mattiat_mapping_2024,rossi_vectorial_2017}.

\subsection{Small angle neutron scattering}\label{AppB}

\begin{figure}[t]%
\centering
\includegraphics[width=1.0\textwidth]{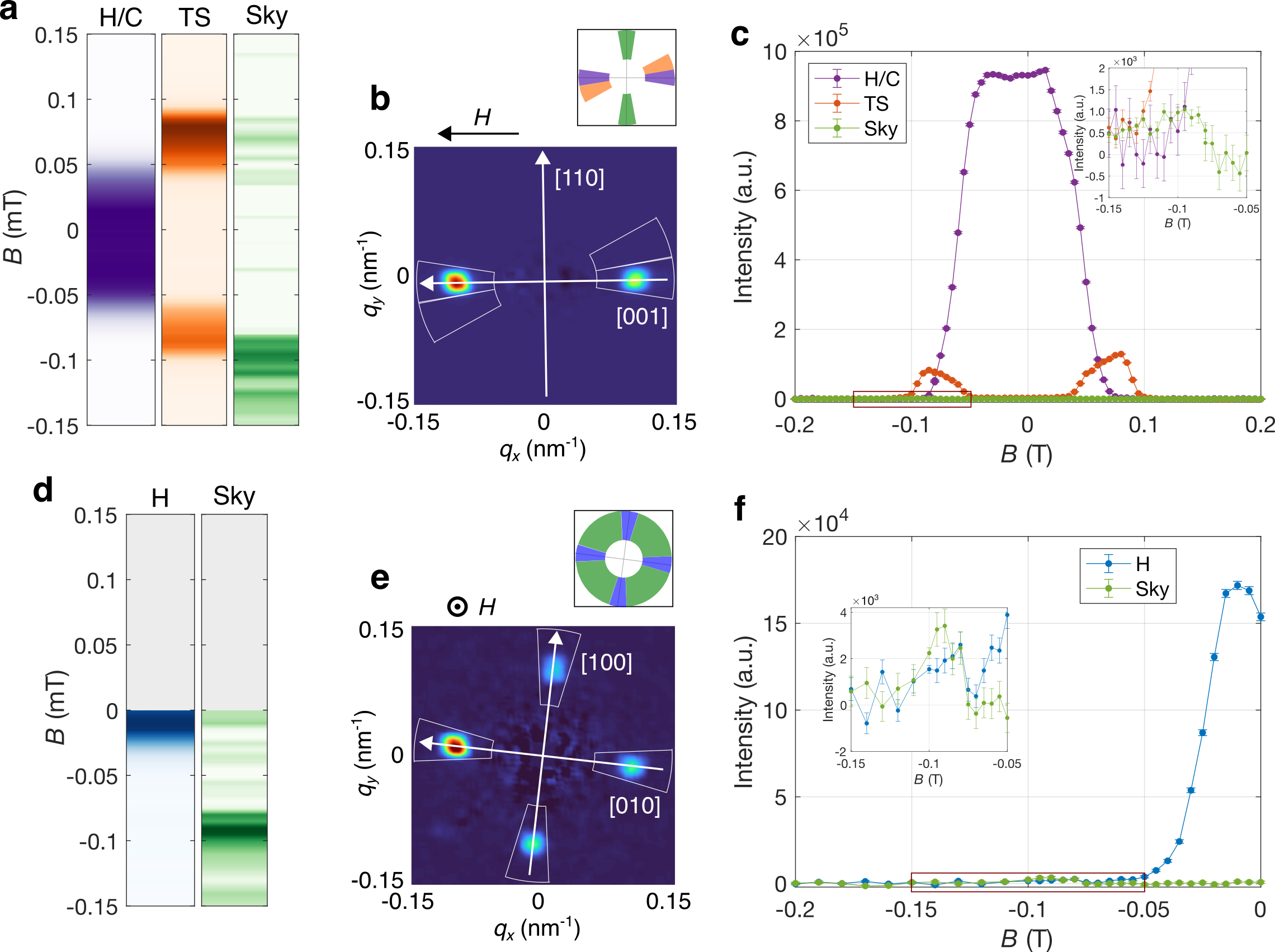}
\caption{Small angle neutron scattering (SANS) of Cu$_2$OSeO$_3$ crystal investigated by SSM and MFM. \textbf{a},~Scattering intensities as a function of applied magnetic field with neutron beam perpendicular to magnetic field for window ranges shown in \textbf{b}. Color scales have been rescaled to the maxima of each type of contrast. \textbf{b},~Corresponding scattering plane in zero field. \textbf{c},~Scattering intensities as a function of applied magnetic field for selected window ranges. \textbf{d},~Scattering intensities as a function of applied magnetic field with neutron beam parallel to the applied magnetic field for window ranges shown in.\textbf{e}. Color scales have been rescaled to the maxima of each type of contrast. \textbf{e}~Corresponding scattering plane in zero field. \textbf{f},~Scattering intensities as a function of applied magnetic field for selected window ranges.}\label{figxx_sans}
\end{figure}

The long-period magnetic structures in the bulk of the crystal are studied using small-angle neutron scattering (SANS). The measurements are performed using the SANS-I instrument at the Swiss Spallation Neutron Source (SINQ), Paul Scherrer Institute (PSI), Switzerland. The crystal is mounted onto a sample stick, and loaded into a horizontal field cryomagnet, which is installed at the neutron beamline. For the SANS measurements, an incident neutron wavelength of 8~\AA ($\Delta\lambda_n/\lambda_n = 10\%$) is selected and collimated over a distance of 18~m before the sample. The scattered neutrons are detected by a two-dimensional multidetector placed 18~m behind the sample. 

The measurement is carried out with the applied field $\mu_0\mathbf{H} \parallel \left[001\right]$ and $\mu_0\mathbf{H} \perp \mathbf{k_i}$, where $\mathbf{k_i}$ is the incident neutron wavevector. In this so-called transverse geometry, SANS signals due to all of the relevant long-period magnetic phases can be observed. Intensity due to H domains with $\mathbf{q_H} \parallel \left<100\right>$, C order with $\mathbf{q_C} \parallel \mu_0 \mathbf{H}$, TS order with $\mathbf{q_{TS}}$ tilted from parallel to $\mu_0\mathbf{H}$, can all be observed in this single instrument geometry. Disordered skyrmion phases can also be studied, since the skyrmion lattice correlations always propagate in a plane perpendicular to $\mu_0\mathbf{H}$. In this transverse field geometry, the associated SANS intensity due to skyrmions forms a ring in the plane perpendicular to $\mu_0\mathbf{H}$, which intersects the two-dimensional detector along the north-south directions. A second measurement with the applied field $\mu_0\mathbf{H} \parallel \left[001\right]$ and $\mu_0\mathbf{H} \parallel \mathbf{k_i}$ is also carried out. In this parallel geometry, only in-plane H domains and skyrmion phases can be observed.

SANS is performed by rotating the cryomagnet and sample stick together over a range of angles that move the various diffraction spots through the Bragg condition at the detector. Detector measurements obtained at each rocking angle are summed together to produce SANS images where all diffraction spots can be observed at once. SANS data are also obtained at fixed rocking angle where intensity from all possible long period phases could be observed simultaneously. Further data are obtained similarly either in the paramagnetic regime at 70~K, or in high-field above magnetic saturation, and used for background subtraction of the low-field / low-temperature data measured below $T_c$. SANS data is analyzed using GRASP software~\cite{dewhurst_graphical_2023}.

We measure the same sample by SANS, which is measured by SSM and MFM, using the same temperature and field protocol. Figs.~\ref{figxx_sans}a and c correspond to measurements performed with the neutron beam perpendicular to the magnetic field applied along the $\left[001\right]$ crystal direction, and Figs.~\ref{figxx_sans}d-f correspond to measurements performed with the neutron beam parallel to the applied magnetic field. Neutron scattering in the parallel configuration is only recorded in reverse field. SANS contrast confirms the presence of the helical, conical, tilted spiral, and low-temperature skyrmion phases.  In the low-field range around zero, SANS measurements indicate the presence of helices along the three $\left<100\right>$-type crystallographic directions, shown in Figs.~\ref{figxx_sans}b and e, both in zero field. 

\begin{figure}[b]%
\centering
\includegraphics[width=0.65\textwidth]{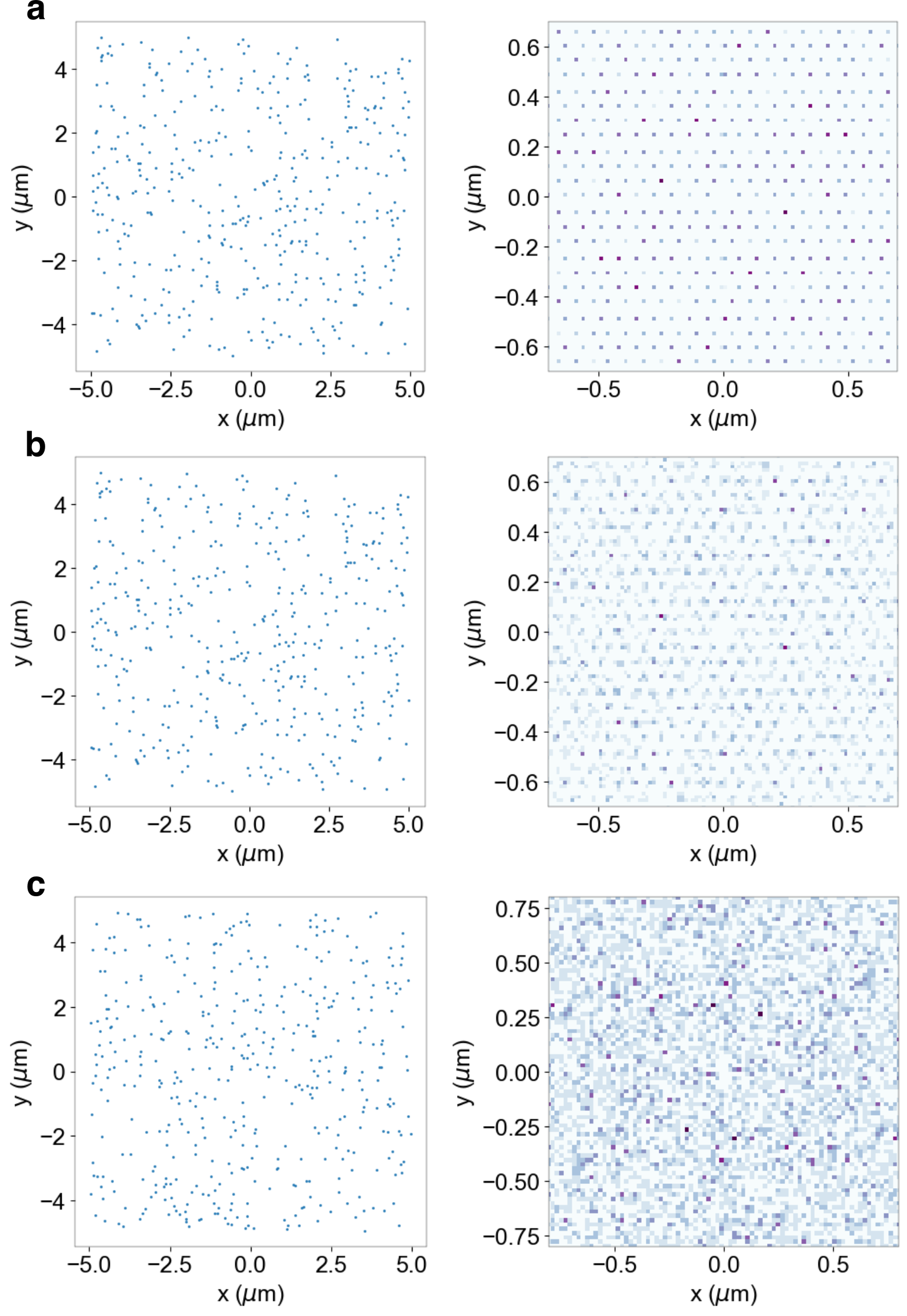}
\caption{Skyrmion lattice simulations with varying degrees of disorder. Each row shows the location of skyrmions in a lattice and the corresponding pair distribution function $G_{p}(\mathbf{r})$, \textbf{a},~with a low skyrmion filling number and perfect lattice periodicity, \textbf{b},~with a low skyrmion filling number and 10\% random deviation from perfect periodicity, and \textbf{c},~with a low skyrmion filling number and 20\% random deviation from perfect periodicity. }\label{figS6_pair}
\end{figure}

\subsection{Simulations of skyrmion pair distribution functions}\label{AppC}

In order to aid the interpretation of our measured skyrmion pair distribution, we simulate a regular skyrmion lattice of 70~nm spacing with an increasing degree of disorder. In particular, we simulate a regular skyrmion lattice with a diluted and random skyrmion site filling. Fig.~\ref{figS6_pair}a shows the locations of randomly distributed skyrmions in a perfectly regular lattice with a similar density as our experimental data, along with its corresponding 2D histogram of the pair distribution function $G_{p}(\mathbf{r})$. We consider the effect of a random deviation in the skyrmion position of 10\% and 20\% from a regular lattice, as shown in Figs.~\ref{figS6_pair}b and c, respectively. Above 20\% deviation, the signatures of translational or orientational order in $G_{p}(\mathbf{r})$ are lost, as in the $G_{p}(\mathbf{r})$ measured at the surface of Cu$_2$OSeO$_3$.



\end{document}